\documentstyle[aps,epsfig]{revtex}

\begin{document}
\title{Scalable Trapped Ion Quantum Computation with a Probabilistic
Ion-Photon Mapping}
\author{L.-M. Duan, B. B. Blinov, D. L. Moehring, C. Monroe \\
FOCUS Center and Department of Physics, University of Michigan,
Ann Arbor, MI 48109 \\}
\maketitle
\begin{abstract}
We propose a method for scaling trapped ions for large-scale
quantum computation and communication based on a probabilistic
ion-photon mapping. Deterministic quantum gates between remotely
located trapped ions can be achieved through detection of
spontaneously-emitted photons, accompanied by the local Coulomb
interaction between neighboring ions. We discuss gate speeds and
tolerance to experimental noise for different probabilistic
entanglement schemes.\\


\end{abstract}

\section{Introduction}

Trapped ions constitute one of the most promising systems for the
implementation of a quantum computer
\cite{0,1,2,3,4,5,6,7,8,9,10,11,10',12}. It appears unlikely that
this system can be scaled by simply adding ions to a single trap,
due to the growing complexity of the vibrational mode spectrum and
the inefficiency of laser cooling of the collective motion of a
large ion crystal to near the ground state. Instead, ion trap
multiplexing can be achieved by either shuttling ions through a
multiply-connected trap structure \cite{10,11}, or coupling
remote ions with a common photon-mediated interaction \cite{0}.
For the latter approach, it is generally believed that the ions
must be enclosed in a high-finesse optical cavity for
deterministic quantum gate operation. This ``strong coupling"
condition means that the coupling rate $g$ between the ion and
the cavity mode should satisfy the requirement
\begin{equation}
g^{2}>>\kappa \gamma _{s}
\end{equation}
where $\kappa $ is the cavity decay rate and $\gamma _{s}$ is the
atomic spontaneous emission rate. Although strong coupling has
been achieved for neutral atoms in recent experiments \cite
{14,14',15}, it remains difficult to do the same with trapped
ions, although there are significant experimental efforts and
achievements \cite{16,16'}. The experimental challenge is that
the required small optical cavity volume can interfere with the
ion trap operation through uncontrolled charges on the dielectric
mirror coatings, and the ion trap electrodes can likewise
interfere with the cavity mode through diffraction.

Alternatively, faithful entanglement can be established between
remote atoms or ions in a probabilistic fashion even if the
strong coupling condition is not satisfied
\cite{17,19,18,20,21,23,22,23'}. Here, the interference of photons
emitted by two remotely-located ions is detected, and a positive
photon count without ``which-path" information as to which ion
emitted the photon will project the two ions into an entangled
state. For this purpose, strong coupling between the ion and the
photon is not essential. Photon loss only affects the success
probability of a positive photon count, and the fidelity of the
entanglement is not reduced. By repeating this kind of entangling
protocol several times, one can ultimately get faithful
entanglement between the ions. We call this a {\it probabilistic}
source of entanglement as the entangling protocol does not
succeed in every trial. With this source of entanglement, one can
construct probabilistic quantum gate operations. However,
probabilistic gates do not in general lead to scalable quantum
computing because of the exponential decrease of the success
probability as the number of quantum gates increases.

In this paper, we show that \textit{remote deterministic} quantum
gates can be constructed for trapped ions from this source of
probabilistic entanglement, when we also allow local deterministic
quantum gates between nearby ions. This provides a method to
scale up the trapped ion system for large scale quantum
computation and communication based on a photon-mediated
interaction without requiring strong coupling between the ion and
photon.

In this scheme, the quantum register consists of a series of ion
pairs (each a logical qubit) that are in different trap regions
separated by arbitrary distances. Within each pair, one ion (the
logic ion) encodes the quantum information and the second ion
(the ancilla) allows the coupling to another ion pair through a
probabilistic entanglement protocol. This probabilistic
entanglement, combined with conventional local motional gates
within each pair, allows for an effective quantum gate between the
remote logical qubits. The resulting remote operation is
deterministic because the probabilistic entanglement operations
can be done off-line, and the failure of an entangling attempt
does not destroy the computational quantum state carried by the
logical ions.

In Sec. II we will show how to achieve deterministic quantum gates
based on the probabilistic entangling operations and how this can
seed a scalable trapped ion quantum computation model. We will
also show how to use the same setup for implementation of quantum
repeaters for achieving scalable long -distance quantum
communication. In Sec. III we will compare two probabilistic
entangling protocols and discuss the gate speeds and tolerance to
noise of each protocol.

\section{Scalable quantum computing and networking from probabilistic
entangling operations}

First, we show how to construct a scalable quantum computation
model for trapped ions by a combination of probabilistic remote
entangling operations and conventional (motional) local gates.
Figure 1 is a schematic of the setup. A series of ion pairs, each
in distinct trap regions, are separated by an arbitrary distance.
Each qubit is represented by a pair of ions, denoted as $i$ and
$i^{\prime }$ respectively for the logic ion and the ancilla ion.
We assume here that the logic ions and the ancilla ions are of
different ion species (or isotopes) so that they can be separately
addressed for laser manipulation and detection through frequency
selection, as has recently been demonstrated in sympathetic
cooling experiments \cite{24,25}. (Of course, they could be the
same species ion if one can achieve spatial separate addressing
through focused laser beams \cite{5}). In this way, the two ions
of a given pair can be tightly confined in a trap with the
ability to operate high-fidelity motional quantum gates between
them.

\begin{figure}[tb]
\epsfig{file=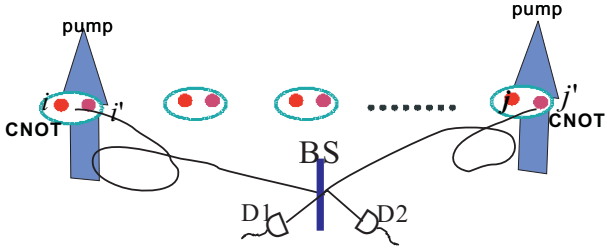,width=8cm} \caption{Schematic illustration
of a quantum computation model based on probabilistic
photon-mediated entanglement between remote ions. The ancilla
ions in different traps are entangled through the probabilistic
protocols described in the next section. Deterministic gates on
remote ions are constructed from the local motional gates and the
probabilistic remote entanglement.} \label{fig1}
\end{figure}

To achieve scalability, we should be able to perform
deterministic quantum gates between two arbitrary logic ions in
different pairs. For this
purpose, we assume that each ancilla ion is connected to a
single-photon detector, possibly through an optical fiber. To
entangle two ancilla ions, say $i^{\prime }$ and $j^{\prime }$,
we pump both with an appropriate resonant laser beam to
excited electronic states. The resulting spontaneously emitted
photons from these two ions are directed to single-photon
detectors for a Bell-type collective measurement. For particular
measurement results, the two ancilla ions $i^{\prime }$ and
$j^{\prime }$
will be projected into a Bell state, which we denote as $%
\left| \Phi _{i^{\prime }j^{\prime }}\right\rangle =\left( \left|
01\right\rangle +\left| 10\right\rangle \right) /\sqrt{2}$ (see
the next section for description of different types of entangling
methods). Each entangling operation
succeeds with probability $p_{s}$ (the probability to register the
appropriate result), so we need to repeat this operation on average
$1/p_{s}$ times for a final successful confirmation of
entanglement, with the total preparation time about
$t_{c}/p_{s}$, where $t_{c}$ is the time for each individual
entangling operation. We assume the logic and ancilla ions are
sufficiently spectrally resolved so that the probabilistic
entangling operation on the ancilla ions does not influence the
logic ions, even if this entangling operation fails.

With the assistance of final Bell state $\left| \Phi _{i^{\prime }j^{\prime
}}\right\rangle $, we can achieve remote quantum controlled-NOT (CNOT) gates
on the logic ions $i$ and $j$. We assume that quantum CNOT gates can be
realized on the local ions $i,i^{\prime }$ and $j,j^{\prime }$ in the same
pairs through conventional means relying on the collective motion of the
ions \cite{1,6,7,12} (note that all the motional gates work even when the two local
ions are of different isotopes or species). These CNOT gates are denoted by $%
C_{ii^{\prime }}$ and $C_{jj^{\prime }}$, where the subscripts
refer to the control and target ions. We can achieve the remote CNOT gate $C_{ij}$ on
the logic ions $i,j$ through a combination of the gates $C_{ii^{\prime }},$ $%
C_{jj^{\prime }}$ and the Bell state $\left| \Phi _{i^{\prime }j^{\prime
}}\right\rangle $. This can be seen by considering the following identity

\begin{eqnarray}
C_{ii^{\prime }}C_{jj^{\prime }}\left( \left| \Psi \right\rangle
_{ij...}\otimes \left| \Phi \right\rangle _{i^{\prime }j^{\prime }}\right)
&=&\left| 0+\right\rangle _{i^{\prime }j^{\prime }}\otimes C_{ij}\left(
\left| \Psi \right\rangle _{ij...}\right) +\left| 0-\right\rangle
_{i^{\prime }j^{\prime }}\otimes \sigma _{i}^{z}C_{ij}\left( \left| \Psi
\right\rangle _{ij...}\right)    \nonumber\\
&&+\left| 1+\right\rangle _{i^{\prime }j^{\prime }}\otimes \sigma
_{j}^{x}C_{ij}\left( \left| \Psi \right\rangle _{ij...}\right) +\left|
1-\right\rangle _{i^{\prime }j^{\prime }}\otimes \left( -\sigma
_{i}^{z}\sigma _{j}^{x}\right) C_{ij}\left( \left| \Psi \right\rangle
_{ij...}\right).
\end{eqnarray}
where $\left| \pm \right\rangle _{j^{\prime }}=\left( \left|
0\right\rangle _{j^{\prime }}\pm \left| 1\right\rangle
_{j^{\prime }}\right) /\sqrt{2}$, and $\left| \Psi \right\rangle
_{ij...}$ denotes the computational state, for which the $i,j$
ions may be entangled with other logic ions. The single qubit
Pauli operators $\sigma_{i}^{z}$ and $\sigma _{j}^{x}$ act on the
corresponding ions $i,j$. The above identity has been implied
previously in different contexts \cite{26,27,28,29}, particularly
in the discussion of the communication complexity of quantum CNOT
gates. The above identity shows that to perform a remote CNOT
gate $C_{ij}$ on the logic ions $i,j$, we can take the following
steps:

\begin{itemize}
\item  Prepare the ancilla ion $i^{\prime }$ and $j^{\prime }$ into the EPR
state $\left| \Phi \right\rangle _{i^{\prime }j^{\prime }}$ using a
probabilistic entangling protocol. Repeat the protocol until it succeeds.

\item  Apply the local motional CNOT gates $C_{ii^{\prime }}$ and $%
C_{jj^{\prime }}$ on the ions $i,i^{\prime }$ and $j,j^{\prime }$ within the
same pairs.

\item  Measure the ancilla ion $i^{\prime }$ in the basis $\left\{ \left|
0\right\rangle _{i^{\prime }},\left| 1\right\rangle _{i^{\prime }}\right\} $
and the ancilla ion $j^{\prime }$ in the basis $\left\{ \left|
+\right\rangle _{j^{\prime }},\left| -\right\rangle _{j^{\prime }}\right\} $.

\item  Apply a single bit rotation $\left\{ I,\sigma _{i}^{z},\sigma
_{j}^{x},-\sigma _{i}^{z}\sigma _{j}^{x}\right\} $ on ion $i$
and/or $j$ if we get the measurement results $\left\{
0+,0-,1+,1-\right\} $, respectively.
\end{itemize}

The resulting remote quantum CNOT gate $C_{ij}$ is deterministic,
even though the seeding entangling operations are probabilistic.
This is because the probabilistic operation can be
repeated off-line until it succeeds. When accompanied by simple
local single-bit rotations, this computation model is therefore scalable,
with no fundamental limit to the number of ion pairs in different traps.
The essential resources are two-ion local motional gates and
remote ion-photon probabilistic entangling operations, both of
which have been demonstrated \cite{2,3,4,5,23'}.

With the same system, we can also realize scalable quantum
networks. The basic problem in quantum networking is to transmit
quantum states over large distances by overcoming the limit
setting by the photon attenuation length. Typically, if one
directly sends a single-photon pulse through an optical channel,
the efficiency (the probability that the photon is not absorbed)
will degrade exponentially with distance due to photon
attenuation. One way to overcome this obstacle is
based on quantum repeaters \cite{30}.
Implementation of quantum repeaters has been proposed in
\cite{19} based on the use of atomic ensembles for storage of
quantum entanglement, and following this scheme some interesting
first-step experiments have been recently reported
\cite{31,32,33}. The quantum repeater can also be realized in the
present context with pairs of ions as discussed above. Figure 2
illustrates schematically the implementation of quantum repeaters
with the paired-ion setup.

With the probabilistic entangling protocol, we can generate
entanglement between two nodes, say $i$ and $k$, and also
$k^{\prime }$ and $j^{\prime }$, each with a communication
distance $L_{0}$ which is smaller or comparable to the photon
attenuation length. The success probability for preparation of
each segment of entanglement is given by $p_{sc}=p_{s}p_{c}$,
where $p_{s}$ is the inherent success probability of the
entangling protocol, and $p_{c}=e^{-\alpha L_{0}}$ is the photon
attenuation in the channel. These two segments of entanglement
can be connected to generate an entangled state between $i$ and
$j^{\prime }$ through a local collective Bell measurement on the
two ions $k$ and $k^{\prime }$ in the same pair. A combination of
a motional CNOT gate and individual ion detections achieves the
desired collective measurement. The preparation time for each
segment of entanglement is $T_{sc}=t_{c}/p_{sc},$ and the time
for establishing entanglement between the next neighboring nodes
$i$ and $j$\ (with a distance $2L_{0}$) is simply estimated by
$T_{2}=2T_{sc}$ if we sequentially prepare each segment of
entanglement. So the time required for establishing entanglement
over $n$ segments with a total communication distance of $nL_{0}$
is estimated by $T_{n}=nT_{sc}=ne^{\alpha L_{0}}\left(
t_{c}/p_{s}\right) $ with the ion-based quantum repeaters. (Here
we neglect local motional gate errors and
ion detection inefficiency, which are typically small compared
with errors from the photon attenuation and the
inherent inefficiency of the entangling protocol). This linear
scaling of the communication time compares favorably with the
exponential scaling law $T_{n}=e^{n\alpha L_{0}}\left(
t_{c}/p_{s}\right) $ for the case of direct communication without
repeaters.

\begin{figure}[tb]
\epsfig{file=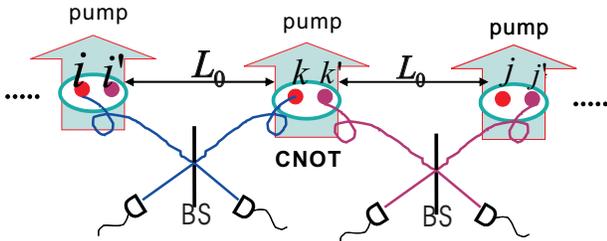,width=8cm} \caption{Schematic illustration
of realization of quantum repeaters with trapped ions based on the
probabilistic detection-induced remote entanglement and the local
Coulomb interaction. } \label{fig2}
\end{figure}

\section{Two types of probabilistic entangling schemes and
their corresponding gate speeds}

We consider two types of probabilistic entangling protocols
\cite{17,20}, denoted as Type I and Type II. Below, we first
introduce the basic ideas of these schemes in the context of the
$^{111}Cd^{+}$ ion \cite{24}.  Then, we compare the merits of
each scheme, with particular attention paid to gate speed and
tolerance to experimental noise.

\subsection{Type I probabilistic entanglement: photon interference}

The Type I protocol was first proposed in Ref. \cite{17} and is
illustrated in Fig. 3. A weak pump pulse is applied to the atomic
transition $\left| 0\right\rangle \rightarrow \left|
e\right\rangle $, which excites the atom to the upper state
$\left| e\right\rangle $ with probability of $p_{e}<<1$.
Spontaneously emitted light from the de-excitation $\left|
e\right\rangle \rightarrow \left| 1\right\rangle $ is collected
within a cone angle $\theta $ (see Fig. 3). The collected light
from the two ions is directed to a beamsplitter (BS) for
interference and then detected through two single photon
detectors D1 and D2. If \textit{one} of the detectors registers a
photon, the two ions will be projected onto the state $\left| \Psi
_{1}\right\rangle =\left| 01\right\rangle $ $+e^{i\varphi }\left|
10\right\rangle $, where the phase $\varphi $ depends on the
difference in path lengths from each ion to the detector.
Conditioned on one detector click, there is a probability $p_{e}$
that both ions spontaneously decay to the state $\left|
1\right\rangle $ with one of the accompanying photons not
registered by the detectors. The inherent infidelity of this
scheme is thus given by $p_{e}$.

\begin{figure}[tb]
\epsfig{file=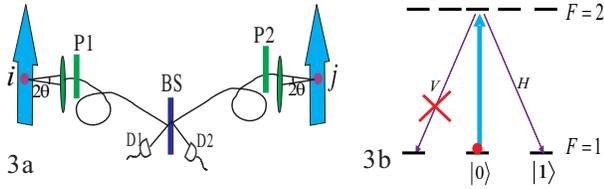,width=8cm} \caption{The type I
probabilistic entangling protocol for two remote atomic qubits.
Fig. 2a shows the atomic level structure from the ground $F=1$
states to the upper $F=2$ states for the $^{111}Cd^{+}$ ion (or
any atomic system with nuclear spin $1/2$ or $3/2$). The
V-polarized light in the collection direction is filtered through
the two polarizers P1 and P2 so it is not relevant for this
protocol. The $\left| 0\right\rangle $ and $\left|
1\right\rangle$ states correspond respectively to the Zeeman
levels  $\left| m=0\right\rangle$ and $(\left| m=+1\right\rangle
+ \left| m=-1\right\rangle )/\sqrt{2}$. The excitation
probability is required to be low and the scheme succeeds when
only one of the detectors fires.} \label{fig3}
\end{figure}

\subsection{Type II probabilistic entanglement: polarization-spin entanglement}

The Type II protocol was first proposed in Ref. \cite{20}. Here,
there are three relevant atomic levels $\left| 0\right\rangle
,\left| 1\right\rangle,\left| a\right\rangle $ in each of two
ions. Figure. 4b shows a configuration of atomic states with an
$F=1$ ground state hyperfine manifold (e.g., within the $S_{1/2}$
ground state of $^{111}Cd^{+}$). The atoms are initially prepared
in state $ \left| a\right\rangle $ and then transferred to states
$\left| 0\right\rangle $ and $\left| 1\right\rangle $ with unit
probability, emitting photons correspondingly either in H or V
polarizations along the direction of collection. The collected
photons traverse a polarization beam splitter (PBS) and are then
detected with two single-photon detectors. The final state of the
two ions will be projected onto $\left|\Psi _{2}\right\rangle
=\left| 01\right\rangle $ $-\left| 10\right\rangle $when
\textit{both detectors} each register a photon. The type II
protocol can also be multiplexed to directly prepare multi-ion
entangled states if the photon collection efficiency is
reasonably high \cite{20}.

\begin{figure}[tb]
\epsfig{file=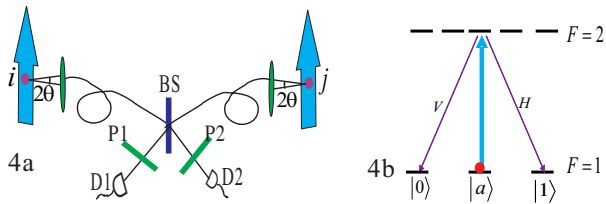,width=8cm} \caption{The type II
probabilistic entangling protocol for two remote ions/atoms. The
polarizer P1 selects the H-polarized light while P2 selects the
V-polarized light. The beam splitter (BS) can be replaced by a
polarization beam splitter (PBS) if P1 and P2 select the
(H+V)-polarized light (rotated to the $45^o$ direction). The
$\left| a\right\rangle $, $\left| 0\right\rangle $ and $\left|
1\right\rangle$ states correspond respectively to the Zeeman
levels  $\left| m=0\right\rangle$, $(\left| m=+1\right\rangle +
\left| m=-1\right\rangle )/\sqrt{2}$, and $(\left|
m=+1\right\rangle - \left| m=-1\right\rangle )/\sqrt{2}$. One
succeeds only when both D1 and D2 register a photon.} \label{fig4}
\end{figure}

\subsection{Comparison between type I and type II probabilistic entanglement}

For the type I protocol, the entanglement success probability, or the probability of detecting a photon is
\begin{equation}
p_{I}=p_{e}p_{c}\eta _{d}/2,
\end{equation}
where $\eta _{d}$ is the detection efficiency, and $p_{c}$ is the
photon collection efficiency, with the form of $p_{c}=3\left(
1-\cos \theta \right) /4$ for the setup schematically shown in
Fig. 3a (in addition to the fraction of the solid angle $\left(
1-\cos \theta \right) /2$, the coefficient $3/2$ comes from the
space integration of the dipole emission pattern). There is an
additional factor of $1/2$ in Eq. (2) since only one polarization
is measured after the beam splitter. So the average entangling
time for this protocol is $T_{I}=t_{c}/p_{I}$, where $t_{c}$ is
the time for one entangling attempt, which is limited only by the
radiative lifetime $t_{e}$ of the upper level $\left|
e\right\rangle $, i.e., $t_{c}>t_{e}\approx 3$ $ns$.

For the type II protocol, the entanglement success probability, or
the probability of registering a single photon from each of the
detectors is
\begin{equation}
p_{II}= p_{c}^{2}\eta _{d}^{2}/8.
\end{equation}
The factor of $8$ appears from the filtering of the two polarizers
and the possibility that two photons go to the same detector.
This probability can be improved by factor of $4$ if we replace
the polarizer by a PBS and add one detector on each side. The
corresponding average entangling time is $T_{II}=t_{c}/p_{II}$.

If the above entangling operations are performed \textit{on-line},
the ancilla ions are entangled during the gate operation, which
is itself the slowest step of the procedure. In this case, the
gate time is then approximately $T_{I}$ or $T_{II}$ for the type
I and type II entangling protocols, respectively. The time
$T_{II}$ is larger than $T_{I}$ if $p_{c}\eta _{d}/4<p_{e}$, and
vice versa. To ensure a reasonable fidelity, the excitation
probability $p_{e}$ is typically about $1\%$, and in the first
experiment with free-space ions \cite{23'}, $ p_{c}\eta _{d}\sim
10^{-3}$ from the limited collection solid angle , so $T_{II}\gg
T_{I}$. Nevertheless, the collection efficiency $p_{c}$ can be
significantly enhanced in future experiments (see the discussion
below), and we might ultimately expect that $T_{I} \sim T_{II}$.
It is also possible to have $T_{II}<T_{I}$ if one puts the ions
into a fairly good cavity even if it is still far from the strong
coupling limit.

The entangling operation can also be done \textit{off-line}. Well
before the desired quantum gate operation, the two ancilla ions
can be entangled through one of the above probabilistic entangling
protocols, and following success this entanglement can be stored
for later quantum gate operations. Here, the potentially slow
off-line entangling operation is not necessarily a limiting issue
for the speed of the subsequent quantum gate operations. The
slowest step of the gate is then the detection of the ancilla
ions, for which the required time can be estimated by $T_{d}\sim
t_{e}/p_{c}\eta _{d}$. With the quantum jump detection method, we
need to register several photons when the ion is at the
``bright'' level, which takes a time of order $t_{e}/(p_{c}\eta
_{d})$.  For instance, with a moderate efficiency $p_{c}\eta
_{d}\sim 10^{-3}$, $T_{d}\sim 10$ $\mu s$ for the $^{111}Cd^{+}$
ion.

Now we discuss the tolerance of the type I and type II
entanglement protocols to relevant experimental noise. For the
type I protocol, the phase $\varphi $ in the entangled state
$\left| \Psi _{1}\right\rangle $ is proportional to $\varphi \sim
\Delta k\Delta x$, where $\Delta k$ is the difference
between the wave vectors for the pumping and collected light, $%
\Delta x$ is the position fluctuation of the ion from its
equilibrium position. The type I protocol thus requires the atom
to be confined within the Lamb-Dicke limit, otherwise the
residual ion motion will randomize the phase $\varphi$ and
consequently degrade the entanglement fidelity.

The type II entanglement protocol is much less sensitive to fluctuations
in the atomic position. This is because the two polarization
components carry the same random phase imposed by the
instantaneous position of the atom. For this common-mode
cancellation of phase fluctuations, the atom must have an
approximately fixed (albeit random) position during the emission
process. This implies that if the ion is not confined within the
Lamb-Dicke limit, the decay rate $1/t_{e}$ of the upper level
$\left| e\right\rangle $ need be significantly larger than the
frequency of any component of ion motion, which is typically the
case. The type-II protocol also exhibits better interferometric
stability \cite{23} for the same
reason that random phases from the two polarization modes cancel
each other. Due to its better noise tolerance, the type II
protocol seems more attractive than the type I protocol, although
the latter could have a higher success probability. An important
seeding step for the type-II protocol has been demonstrated in a
recent experiment \cite{23'}, where entanglement has been
directly observed for the first time between a stationary ion spin
qubit and a flying photon polarization qubit.

Finally, we consider possibilities for improving the collection
efficiency of the photons emitted by the ions. One way is to position
the ion in an optical cavity, whereby the effective collection efficiency
can be improved by a factor of the cavity finesse compared with
the free space case with the same collection solid angle. Even
for cavity mirrors with a moderate finesse, the success
probability, and thus the gate time, for either protocol above
can be significantly improved (particular for the type II
protocol, as $T_{II}$ scales quadratically with $p_{c}$).
Alternatively, the emitted light can be collected by a
single-mode optical fiber. This not only filters unwanted spatial
modes so that the emission from multiple atoms can be easily
mode-matched, but also the collection efficiency could be
improved with near-field engineering of the fiber tips. Here, the
angular dependence of the emitted photon polarization may cause a
reduced amount of entanglement. However, clever linear optical
transformations may allow high-fidelity entangled states to be
recovered, with some tradeoff in efficiency \cite {hhh}.

In summary, we have proposed a new method to scale up the ion trap system for
large scale quantum computing or networking, based on a probabilistic ion-photon
mapping. Remarkably, the photon-mediated interaction does not require strong
coupling between the ion and the photon.  The protocol can also be highly
tolerant of relevant experimental imperfections.

\end{document}